# Structural, electronic and magnetic properties of η carbides ($Fe_3W_3C$, $Fe_6W_6C$, $Co_3W_3C$ and $Co_6W_6C$) from first principles calculations

D.V. Suetin, I.R. Shein, A.L. Ivanovskii *

*Institute of Solid State Chemistry, Ural Branch of the Russian Academy of Sciences, Ekaterinburg GSP-145, 620041 Russia*

**Abstract**

First-principles FLAPW-GGA calculations have been performed with the purpose to determine the peculiarities of the structural, electronic, magnetic properties and stability for a family of related η carbides $M_3W_3C$ and $M_6W_6C$ (where M= Fe and Co). The geometries of all phases were optimized and their structural parameters, theoretical density, cohesive and formation energies, total and partial densities of states, atomic magnetic moments have been obtained and analyzed in comparison with available theoretical and experimental data.

*Keywords:* η carbides $M_3W_3C$, $M_6W_6C$ (M= Fe, Co); electronic, cohesive, structural, magnetic properties; FLAPW-GGA calculations

## 1. Introduction

Tungsten carbide WC attracts much attention thanks to its unique physical and chemical properties such as extreme hardness, high melting point, chemical inertness, interesting catalytic behavior *etc*, and belongs to the most promising engineering materials with a wide range of industrial applications, for example in high temperature tools and devices: high speed tools, extrusion dies, rollers, drills, etc. Recent applications include their usage in catalysis industries and as aerospace coatings [1-8].

Simultaneously, significant attention is given to crystalline and nano-sized WC-based alloys and composites comprising other transition metals (WC-M, where M are *d* metals), which are suitable for many technological applications, see [8-12].

Today, a series of various materials has been prepared in the WC-M systems. One extensive group of such materials includes the above-mentioned composites (the so-called cemented carbides), i.e. heterogeneous WC-M systems which consist of grains of tungsten carbide glued with a binder metal M to combine the hardness of the carbide and the toughness of the metal, see [8].

Another important group consists of double (M-W) carbides, which adopt individual crystal structures and properties. This group includes the so-called η carbides such as $Fe_3W_3C$ or $Co_6W_6C$ [1-3,8]. These phases may arise in heterogeneous composites in the interface region between WC and transition metals (or their alloys) [13-21] or may be prepared using special synthetic routes - for example, by mechanical alloying [22-24].

Today among the tungsten-containing η carbides several related phases with various M/W content are obtained (for example $Co_2W_4C$ and $Co_4W_2C$, see [8]), but usually η carbides with stoichiometry M/W=1, namely $M_3W_3C$ and $M_6W_6C$, are found [1-3,8-24]. These η carbides possess interesting properties. For example, the bulk modulus for $Co_6W_6C$ was found [25] to be 462 GPa, *i.e.* higher than that for diamond (~ 444 GPa, see [26]) and WC (~ 421 [27]). The Vickers microhardness ($H_V$) measurements for $Fe_3W_3C$ and $Fe_6W_6C$ show that they are hard phases with $H_V$ ~ 15.6 GPa [24]. These phases are of high technological importance: for example, the properties of widely used WC/M composite materials as well as tungsten-containing steels and heavy alloys are substantially determined by the formation of η carbides as secondary phases [8,13-21].

On the other hand, extensive theoretical studies have been performed for the electronic structure, stability and physical properties of binary tungsten carbides [28-36], whereas the data about the fundamental electronic properties of η carbides $M_3W_3C$ and $M_6W_6C$ are practically absent: to our knowledge, only in one earlier work [37] the electronic spectra for the non-magnetic $Fe_3W_3C$ and $Fe_6W_6C$ are discussed based on the cluster model.

In the present work, using the full-potential linearized augmented plane waves (FLAPW) method within the generalized gradient approximation (GGA) for the exchange-correlation potential we explore for the first time the structural, electronic, magnetic and cohesive characteristics of the double tungsten-containing η carbides $Fe_3W_3C$, $Fe_6W_6C$, $Co_3W_3C$ and $Co_6W_6C$.

The choice of these four η phases allows us to compare the above properties of the related double carbides depending on the metal/carbon ratio ((M,W)/C=6 *versus* (M,W)/C=12), *i.e.* ($Fe_3W_3C$, $Co_3W_3C$) ↔ ($Fe_6W_6C$, $Co_6W_6C$), as well as on the transition 3$d$ metal type (Fe *versus* Co, *i.e.* $Fe_3W_3C$ ↔$Co_3W_3C$ and $Fe_6W_6C$↔ $Co_6W_6C$).

As a result, the optimized structural parameters, theoretical density, cohesive and formation energies, total and partial densities of states (DOS) and atomic magnetic moments for the η carbides $M_3W_3C$ and $M_6W_6C$ (where M = Fe and Co) have been obtained and analyzed in comparison with available theoretical and experimental data.

## 2. Models and method of calculations

According to available crystallographic data [1,3,38-40], all the examined η carbides $Fe_3W_3C$, $Fe_6W_6C$, $Co_3W_3C$ and $Co_6W_6C$ adopt the cubic symmetry with the space group *Fd*3*m* (No 227) and $Z$ = 16 (for $Fe_3W_3C$ and $Co_3W_3C$) and $Z$ = 8 (for $Fe_6W_6C$ and $Co_6W_6C$). In both crystal structures ($M_3W_3C$ and $M_6W_6C$, where M = Fe or Co), the tungsten atoms occupy the 48$f$ sites; Fe and Co are placed in two non-equivalent 32$e$ ($M_1$) and 16$d$ ($M_2$) sites, whereas the carbon is located in the 16$c$ (for $M_3W_3C$) or the 8$a$ sites – for $M_6W_6C$, Table 1.

The ideal structure of the cubic η carbides $M_3W_3C$ is quite complicated and consists of eight regular octahedra of tungsten atoms centered in a diamond cubic lattice and eight regular tetrahedra of Fe(Co) atoms centered in the second diamond cubic lattice that interpenetrates the first through the 1/2, 1/2, 1/2 unit cell translation. Sixteen additional Fe(Co) atoms are tetrahedrally coordinated around the Fe(Co) tetrahedra and sixteen carbon atoms surround the tungsten octahedra in tetrahedral coordination.

The only difference between the above $M_3W_3C$ and $M_6W_6C$ is that these phases contain 16 and 8 carbon atoms (per cell), respectively, and for $M_6W_6C$, the carbon atoms occupy the 8*a* sites in the octahedral coordination [1,38,40], see Fig. 1.

Our band-structure calculations for all η carbides $M_3W_3C$ and $M_6W_6C$ were performed within the full-potential method with mixed basis APW+lo (LAPW) implemented in the WIEN2k suite of programs [41]. The generalized gradient correction (GGA) to exchange-correlation potential of Perdew, Burke and Ernzerhof [42] was used. The electronic configurations were taken to be [Xe] $6s^2 5d^4$ for W, [Ar] $4s^2 3d^6$ for Fe, [Ar] $4s^2 3d^7$ for Co and [He] $2s^2 2p^2$ for carbon. Here, the noble gas cores were distinguished from the sub-shells of valence electrons. The basis set inside each muffin-tin (MT) sphere was split into core and valence subsets. The core states were treated within the spherical part of the potential only, and were assumed to have a spherically symmetric charge density in MT spheres. The valence part was treated with the potential expanded into spherical harmonics to $l = 4$. The valence wave functions inside the spheres were expanded to $l = 12$. The plane-wave expansion with $R_{MT} \times K_{MAX}$ was equal to 7, and $k$ sampling with 5×5×5 $k$-points mesh in the Brillouin zone was used. Relativistic effects were taken into account within the scalar-relativistic approximation. The MT atomic radii were 1.97 a.u. for W, Fe, Co, and 1.75 a.u. for carbon.

The self-consistent calculations were considered to have converged when the difference in the total energy of the crystal did not exceed 0.01 mRy as calculated at consecutive steps. In this way we used the optimization regime as was described in the original version WIEN2k [41]; this means minimization of the total energy by variation of the lattice parameters (*a*) and minimization of the atomic forces (< 1 mRy/a.u.). The density of states (DOS) was obtained using the modified tetrahedron method [43]. Finally, to examine the magnetic properties of the η carbides, our calculations were carried out for non-magnetic and magnetic variants. In the latter case the ferromagnetic ordering (FM) was assumed.

## 3. Results and discussion

### 3.1. Structural properties and density

At the first step, the equilibrium lattice constants ($a$) and cell volumes ($V$) for the ideal stoichiometric η carbides $Fe_3W_3C$, $Fe_6W_6C$, $Co_3W_3C$ and $Co_6W_6C$ were calculated. The results are presented in Table 2. As can be seen, $a(Fe_3W_3C) > a(Co_3W_3C)$ and $a(Fe_6W_6C) > a(Co_6W_6C)$; this result can be easily explained by the atomic radii of 3$d$ metals: R(Fe) = 1.26 Å > R(Co) = 1.25 Å. On the other hand, comparison of $M_3W_3C$ and $M_6W_6C$ phases shows that $a(Fe_3W_3C) > a(Fe_6W_6C)$ and $a(Co_3W_3C) > a(Co_6W_6C)$. These results agree well with the experimental data (Table 2); the deviation of our results from the experimental data does not exceed 0.2-0.8% and is connected with the use of the GGA formalism.

Then the calculated cell volumes were used to estimate the theoretical density ($\rho^{theor}$) of the considered phases. The data obtained showed that the density of these materials decreased in the following sequence: $\rho^{theor}(Co_6W_6C) > \rho^{theor}(Fe_6W_6C) > \rho^{theor}(Co_3W_3C) > \rho^{theor}(Fe_3W_3C)$. Thus, the Co-containing phases have a higher density than Fe-containing phases (with the same (M,W)/C content) as well as the phases with a smaller content of carbon (i.e. $\rho^{theor}(M_6W_6C) > \rho^{theor}(M_3W_3C)$). At the same time, all four η carbides are lighter than the binary tungsten carbide – the hexagonal WC ($\rho^{theor}$ = 15.395 g/cm$^3$ [36] and $\rho^{exp}$ =15.5 ÷ 15.7 g/cm$^3$ [44]).

### 3.2. Formation energies

To provide an insight into the fundamental aspects of phase equilibrium in the (Fe,Co)-W-C systems and determine the relative stability of the examined η carbides $M_3W_3C$ and $M_6W_6C$, their formation energies ($E_{form}$) were estimated on the basis of our total energy calculations.

Let us note that the stability of the known binary phases forming in the systems (Fe,Co)-C and W-C is quite different: the tungsten carbide (hexagonal WC) is a very stable compound, while $(Fe,Co)_3C$ carbides are thermodynamically unstable and are synthesized in a non-equilibrium process [1-3]. The theoretical estimations of the formation energies for WC and $(Fe,Co)_3C$ confirm this situation very well: the obtained $E_{form}$ for WC and $(Fe,Co)_3C$ (calculated as $E_{form}(WC) = E_{tot}(WC) - \{E_{tot}(W) + E_{tot}(C^g)\}$ and $E_{form}((Fe,Co)_3C) = E_{tot}((Fe,Co)_3C) - \{3E_{tot}(Fe,Co) + E_{tot}(C^g)\}$, where $E_{tot}(W,Fe,Co)$ and $E_{tot}(C^g)$ are the total energies for metallic W, Fe, Co and Bernal-type graphite with …*ABAB*…interlayer stacking) are: $E_{form}(WC) = -0.17$ eV/atom [36], $E_{form}(Fe_3C) = +0.22$ eV/atom and $E_{form}(Co_3C) = +0.49$ eV/atom [45].

Here the same approach was used for the estimation of the relative stability of the η carbides – for the reactions with participation of simple substances (metallic Fe, Co, W and graphite). In this way, the formation of $Fe_6W_6C$, for example, was considered for the formal reaction $6Fe + 6W + C \rightarrow Fe_6W_6C$. Then, the formation energy of $Fe_6W_6C$ was defined as: $E_{form}(Fe_6W_6C) = E_{tot}(Fe_6W_6C) - \{6E_{tot}(Fe) + 6E_{tot}(W) + E_{tot}(C^g)\}$. Note that within this definition the negative $E_{form}$ indicates that it is energetically favorable for given reagents to mix and form stable η carbides, and *vice versa.*

The results obtained (Table 3) show that synthesis of the double carbides from simple substances (metallic W, Fe(Co) and graphite) is favorable, their negative $E_{form}$ increasing in the following sequence: $E_{form}(Co_3W_3C) > E_{form}(Fe_3W_3C) > E_{form}(Co_6W_6C) > E_{form}(Fe_6W_6C)$. Thus, among the examined η carbides the most thermodynamically stable compound should be $Fe_6W_6C$ and the most unstable – $Co_3W_3C$. At the same time, all η carbides are less stable than the binary tungsten carbide – the hexagonal WC, Table 3. This conclusion is supported by the calculated values of the cohesive energy ($E_{coh}$, defined as the difference between total energies of η carbides (or WC) and free W, M and carbon atoms). Our results (Table 3) show that $E_{coh}$ for all η carbides are smaller than that for the pure carbide h-WC.

This agrees with the experiments [15,18,46,47], in which at the first stage starting products containing mainly WC and Fe (Co) are obtained in the ternary (Fe,Co)-W-C systems, and after the activation procedures (different ball-milling processes, heat treatment *etc*.) double η carbides begin to form as intermediate phases.

We have also examined the formation energies of the η carbides for some possible preparation routes, namely with participation of the *mono*-carbide h-WC and the most stable polymorph of tge tungsten *semi*-carbide ε-$W_2C$ [36]. Here the formal reactions 3Fe + 3 h-WC → $Fe_3W_3C$ + 2$C^g$ and 3Co + 3/2 ε-$W_2C$ → $Co_3W_3C$ + 1/2 $C^g$ were treated; and the corresponding formation energies of $Fe_3W_3C$ and $Co_3W_3C$ were defined as: $E_{form}(Fe_3W_3C)$ = $E_{tot}(Fe_3W_3C)$ – {3$E_{tot}$(Fe) + 3$E_{tot}$(WC) – 2$E_{tot}(C^g)$} and $E_{form}(Co_3W_3C)$ = $E_{tot}(Co_3W_3C)$ – {3$E_{tot}$(Co) + 3/2$E_{tot}$(ε-$W_2C$) – ½ $E_{tot}(C^g)$}, respectively.

We found (Table 4) that the formation of the η carbides in reactions with the participation of the tungsten *semi*-carbide is favorable – in contrast to the reactions with the participation of the *mono*-carbide, where the values of $E_{form}$ are positive. Probably, this can be explained taking into account the above mentioned high stability of the binary h-WC, when the replacement of W atoms by Fe or Co atoms is energetically unfavorable. Thus, thermodynamic factors such as temperature and pressure should be taken into account while comparing these data with experiments.

*3.3. Electronic structure and magnetic properties.*

In our recent work [48] devoted to a series of h-WC based solid solutions $W_{1-x}M_xC$, where M are 3*d* metals: Sc, Ti….Ni and Cu, it was established that all the considered carbides $W_{1-x}M_xC$ are non-magnetic, except $W_{1-x}Co_xC$, for which the magnetic state is more energetically favorable - owing to spin polarization of Co 3$d$ states, and the corresponding atomic magnetic moment on cobalt MM~ 0.84 $\mu_B$.

Thus, at the first step we have examined the possibility of magnetization of the considered η carbides. The results of our calculations for NM and FM states indicate that unlike the above solid solutions $W_{1-x}(Fe,Co)_xC$, the ground state for $Co_3W_3C$ and $Co_6W_6C$ is non-magnetic, whereas the magnetic behavior for Fe-containing η carbides is preferable: the energy preference of the ferromagnetic state *versus* NM is about 0.001 eV per atom for $Fe_3W_3C$ and about 0.002 eV per atom for $Fe_6W_6C$.

Let us consider the main peculiarities of the electronic properties for the η carbides using NM $Co_3W_3C$ and $Co_6W_6C$ as an example. Figures 2 and 3 show the band structures, total and partial density of states (DOS) for $Co_3W_3C$ and $Co_6W_6C$. These materials with related crystal structures exhibit some common features of their electronic spectra. Namely, in both carbides, the valence bands (VB) extend from -13.7 eV to the Fermi level $E_F$ = 0 eV (for $Co_3W_3C$) and from -12.6 eV to $E_F$ (for $Co_6W_6C$) and are derived basically from the W 5$d$, Co 3$d$ and C 2$s$,2$p$ states. The C 2$s$ states (peak A Fig. 2) are situated in the region from -13.7 eV to -12 eV below the Fermi level, whereas W and Co states play a relatively minor role in this interval. The occupied (W+Co) $d$ and C 2$p$ bands are localized in the region from -8.2 to $E_F$ and from -8.4 to $E_F$ for $Co_3W_3C$ and $Co_6W_6C$, respectively; these states are separated from the C 2$s$ states by gaps of about 3.7-3.6 eV.

The W 5$d$ bands are much more extended as compared with the Co 3$d$ bands. As a result, the bottom of the common (W+Co) $d$ band (peak B Fig. 2) is composed mostly of the W 5$d$ states hybridized with the C 2$p$ states, forming covalent W-C bonds.

The top of the VB and the bottom of the conduction band (peaks C and D Fig. 2) are of a mixed (W+Co) $d$ character; and therefore the η carbides belong to the metallic-like materials. The calculated band structure parameters (band width) for all η carbides are presented in Table 5 in comparison with h-WC. It is seen that the width of the VB for the considered η phases varies maximum by 0.23 eV and is equal to 8.2 – 8.4 eV, Table 5.

For all η carbides the Fermi level lies in the region of mixed W+(Fe,Co) $d$ bands with a high density of states at the Fermi level $N(E_F)$, Table 6, and in the NM state the Fe-containing phases have higher $N(E_F)$ than Co-containing phases ($N(E_F)(Fe_3W_3C) > N(E_F)(Co_3W_3C)$ and $N(E_F)(Fe_6W_6C) > N(E_F)(Co_6W_6C)$). This is a qualitative factor of non-magnetic instability for the Fe-containing η carbides and leads to their magnetization. Besides, as going from $M_3W_3C$ to $M_6W_6C$ (*i.e.* with reduction of carbon content) the values of $N(E_F)$ grow.

The calculated values of partial DOSs at the Fermi level, Table 6, show that the W $5d$ and (Fe,Co) $3d$ states give comparable contributions, whereas the contributions of the C $2p$ states are much smaller. Besides, the contributions from the $3d$ atoms placed in the $32e$ sites ($M_1$) are noticeably higher than for the $M_2$ atoms placed in the $16d$ sites. Let us note that this picture is quite different from the h-WC, where the Fermi level is situated close to the DOS minimum, with the main contribution of the W $5d$ states [36].

Finally, returning to magnetic carbides $Fe_3W_3C$ and $Fe_6W_6C$ (Fig. 3, Table 7), we shall note that magnetism originates mainly from spin polarization of the Fe $3d$ states, whereas induced magnetization on tungsten and carbon atoms is small; besides, the calculated local magnetic moments for $Fe_1$ and $Fe_2$ atoms are quite different (Table 7) and depend on the type of their atomic position.

## 4. Conclusions

In conclusion, the first-principles FLAPW-GGA band structure calculations have been performed to explore the structural, cohesive, electronic, and magnetic properties of the so-called double η carbides $M_3W_3C$ and $M_6W_6C$ (where M = Fe and Co), which greatly affect the properties of widely used WC/M composites, as well as tungsten-containing steels and heavy alloys. The results obtained are summarized as follows:

(i) The calculations of the equilibrium lattice constants ($a$) showed that $a(Fe_3W_3C) > a(Co_3W_3C)$ and $a(Fe_6W_6C) > a(Co_6W_6C)$; on the other hand, for the $M_3W_3C$ and $M_6W_6C$ phases $a(Fe_3W_3C) > a(Fe_6W_6C)$ and $a(Co_3W_3C) > a(Co_6W_6C)$. The density of these materials decreases in the following sequence: $\rho(Co_6W_6C) > \rho(Fe_6W_6C) > \rho(Co_3W_3C) > \rho(Fe_3W_3C)$, i.e. the Co-containing phases have higher density than the Fe-containing phases (with the same (M,W)/C content) as well as the phases with a smaller content of carbon (i.e. $\rho(M_6W_6C) > \rho(M_3W_3C)$). At the same time, all η carbides are lighter than the hexagonal binary tungsten carbide WC.

(ii) The estimations of formation energies ($E_{form}$) show that synthesis of the double carbides from simple substances (metallic W, Al and graphite) is favorable, and their negative $E_{form}$ increase in the following sequence: $E_{form}(Co_3W_3C) > E_{form}(Fe_3W_3C) > E_{form}(Co_6W_6C) > E_{form}(Fe_6W_6C)$. Thus, among the examined η carbides the most thermodynamically stable material should be $Fe_6W_6C$, the most unstable – $Co_3W_3C$. At the same time, all η carbides are less stable than the hexagonal binary tungsten carbide WC. Additional estimations of $E_{form}$ for some possible preparation routes reveal that synthesis of the η carbides in reactions with the participation of the tungsten *semi*-carbide ε-$W_2C$ is favorable – in contrast to the reactions with the participation of the h-WC *mono*-carbide.

(iii) These materials with the related crystal structures have some common features of the electronic spectra, namely, the valence bands of all η carbides are derived basically from the W 5*d*, (Fe,Co) 3*d* and C 2*s*,2*p* states. The W 5*d* bands hybridized with the C 2*p* states (forming covalent W-C bonds) are more extended as compared with the (Fe,Co) 3*d* bands, for which overlapping with the C 2*p* states is weaker. For all η carbides the Fermi level lies in the region of mixed W+(Fe,Co) *d* bands with a high density of states at the Fermi level N($E_F$). The calculated values of partial DOSs at the Fermi level are indicative of comparable contributions from the W 5*d* and (Fe,Co) 3*d* states, whereas the contributions of the C 2*p* states are much smaller. The contributions from the 3*d* atoms placed in the 32$e$ sites (M$_1$) are noticeably higher than for the M$_2$ atoms placed in the 16$d$ sites. On the whole, the Fe-containing phases have higher values of N($E_F$) than the Co-containing phases, which is a qualitative factor of non-magnetic instability for the Fe-containing η carbides leading to their magnetization.

(vi) Magnetism of Fe$_3$W$_3$C and Fe$_6$W$_6$C originates mainly from spin polarization of the Fe 3$d$ states, whereas induced magnetization on the tungsten and carbon atoms is quite small. Besides, the calculated local magnetic moments for Fe$_1$ and Fe$_2$ atoms are quite different and depend on the type of their atomic coordination.

**Acknowledgements.** The work was supported by RFBR, grant No. 08-08-00034.

Table 1.

Atomic positions for the cubic η carbides $M_3W_3C$ and $M_6W_6C$, where M = Fe or Co [1,38-40].

| η carbides | W | $M_1$ | $M_2$ | C |
|---|---|---|---|---|
| $M_3W_3C$ | 48 *f* ($x_1$;0.125;0.125) | 32 *e* ($x_2$; $x_2$; $x_2$) | 16 *d* (0.5;0.5;0.5) | 16 *c* (0;0;0) |
| $M_6W_6C$ | 48 *f* ($x_1$;0.125;0.125) | 32 *e* ($x_2$; $x_2$; $x_2$) | 16 *d* (0.5; 0.5; 0.5) | 8 *a* (0.125;0.125;0.125) |

Table 2.

Optimized lattice parameters (*a*, in nm), cell volumes (V, in nm$^3$), internal atomic positions ($x_1$ and $x_2$, see Table 1) and theoretical density ($\rho^{theor}$, in g/cm$^3$) for the cubic η carbides $M_3W_3C$ and $M_6W_6C$, (where M = Fe or Co) in comparison with available experimental data [38-40].

| η carbides | *a* | V | $x_1$ | $x_2$ | ρ |
|---|---|---|---|---|---|
| $Co_3W_3C$ | 1.1023 (1.1112 [38]) | 0.3349 | 0.3314 | 0.7085 | 14.685 |
| $Fe_3W_3C$ | 1.1032 (1.1087 [39]) | 0.3357 | 0.3278 (0.3228 [39]) | 0.7024 (0.7047 [39]) | 14.464 |
| $Co_6W_6C$ | 1.0877 (1.0897 [38]) | 0.3217 | 0.3248 | 0.7080 | 15.163 |
| $Fe_6W_6C$ | 1.0900 (1.0934 [40]) | 0.3238 | 0.3233 | 0.7035 | 14.873 |

Table 3.
Calculated formation energies ($E_{form}$, eV/atom) and cohesive energies ($E_{coh}$, eV/atom) for the cubic η carbides $M_3W_3C$ and $M_6W_6C$, where M = Fe or Co, in comparison with the hexagonal tungsten carbide WC [34,36].

| η carbides | $Co_3W_3C$ | $Fe_3W_3C$ | $Co_6W_6C$ | $Fe_6W_6C$ | h-WC |
|---|---|---|---|---|---|
| - $E_{form}$ | 0.08 | 0.12 | 0.14 | 0.16 | 0.17 [36] |
| - $E_{coh}$ | 8.91 | 8.70 | 8.97 | 8.72 | 10.69 [34] |

Table 4.
Calculated formation energies ($E_{form}$, eV/atom) for the cubic η carbides $M_3W_3C$ and $M_6W_6C$, where M = Fe or Co, for some possible synthetic routes with participation of the binary tungsten *mono-* carbide (h-WC) and *semi-* carbide (ε-$W_2C$).

| Formal reaction | $E_{form}$ |
|---|---|
| 3Fe + 3 h-WC → $Fe_3W_3C$ + 2$C^g$ | 0.03 |
| 6Fe + 6 h-WC → $Fe_6W_6C$ + 5$C^g$ | 0.00 |
| 3Fe + 3/2 ε-$W_2C$ → $Fe_3W_3C$ + 1/2 $C^g$ | -0.11 |
| 6Fe + 3 ε-$W_2C$ → $Fe_6W_6C$ + 2$C^g$ | -0.15 |
| 3Co + 3 h-WC → $Co_3W_3C$ + 2$C^g$ | 0.06 |
| 6Co + 6 h-WC → $Co_6W_6C$ + 5$C^g$ | 0.01 |
| 3Co + 3/2 ε-$W_2C$ → $Co_3W_3C$ + 1/2 $C^g$ | -0.07 |
| 6Co + 3 ε-$W_2C$ → $Co_6W_6C$ + 2$C^g$ | -0.13 |

Table 5.
Calculated band structure parameters (band width, in eV) for the non-magnetic η carbides $M_3W_3C$ and $M_6W_6C$, where M = Fe or Co, in comparison with h-WC [36].

| system | Band types | | | |
|---|---|---|---|---|
| | Common band (C 2$s$ ÷ $E_F$) | C 2$s$ band | gap C 2$s$ ÷ C 2$p$ + (W+M) $d$ | Valence band C 2$p$ + (W=M) $d$ (to $E_F$) |
| $Co_3W_3C$ | 13.66 | 1.81 | 3.66 | 8.19 |
| $Fe_3W_3C$ | 13.85 | 1.88 | 3.72 | 8.25 |
| $Co_6W_6C$ | 12.64 | 0.59 | 3.63 | 8.42 |
| $Fe_6W_6C$ | 12.68 | 0.57 | 3.84 | 8.27 |
| WC [36] | 14.51 | 4.22 | 1.95 | 8.34 |

Table 6.
Total N($E_F$) and partial N$^l$($E_F$) densities of states at the Fermi level (in states/eV·atom) for the non-magnetic η carbides $M_3W_3C$ and $M_6W_6C$, where M = Fe or Co, in comparison with h-WC [36].

| system | N($E_F$) | N$^{W\,5d}$($E_F$) | N$^{M\,3d}_1$($E_F$) | N$^{M\,3d}_2$($E_F$) | N$^{C\,2p}$($E_F$) |
|---|---|---|---|---|---|
| $Co_3W_3C$ | 2.253 | 0.426 | 0.853 | 0.453 | 0.044 |
| $Fe_3W_3C$ | 2.506 | 0.488 | 0.890 | 0.584 | 0.035 |
| $Co_6W_6C$ | 2.336 | 0.421 | 0.986 | 0.454 | 0.011 |
| $Fe_6W_6C$ | 2.564 | 0.438 | 1.052 | 0.568 | 0.011 |
| WC [36] | 0.228 | 0.099 | - | - | 0.029 |

Table 7.
Spin-resolved total and partial densities of states at the Fermi level (N($E_F$) and N$^l$($E_F$) in states/eV·atom·spin) and local magnetic moments μ(in μ$_B$) for the non-equivalent $Fe_1$ and $Fe_2$ atoms (placed in 32$e$ and 16$d$ sites, respectively) for the magnetic η carbides $Fe_3W_3C$ and $Fe_6W_6C$.

| carbide | N($E_F$) | N$^{W\,5d}$($E_F$) | N$^{M\,3d}_1$($E_F$) | N$^{M\,3d}_2$($E_F$) | N$^{C\,2p}$($E_F$) | μ($Fe_1$) | μ($Fe_2$) |
|---|---|---|---|---|---|---|---|
| $Fe_3W_3C$ | 0.960(↑) | 0.193(↑) | 0.311(↑) | 0.225(↑) | 0.019(↑) | -0.11 | 0.75 |
|  | 0.558(↓) | 0.092(↓) | 0.211(↓) | 0.141(↓) | 0.005(↓) |  |  |
| $Fe_6W_6C$ | 0.851(↑) | 0.152(↑) | 0.413(↑) | 0.115(↑) | 0.003(↑) | 0.25 | 0.58 |
|  | 0.759(↓) | 0.132(↓) | 0.318(↓) | 0.159(↓) | 0.003(↓) |  |  |

**FIGURES**

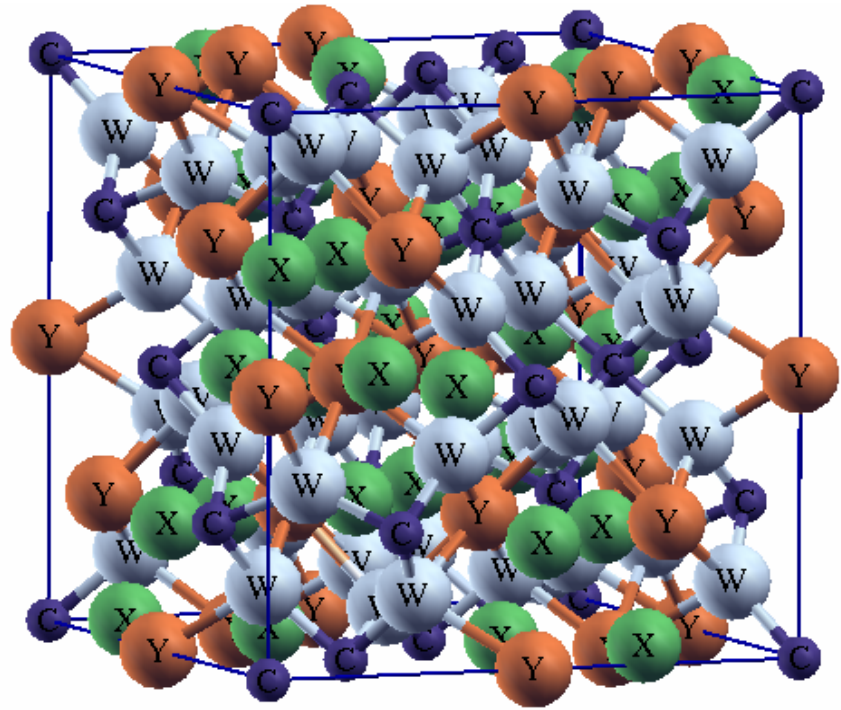

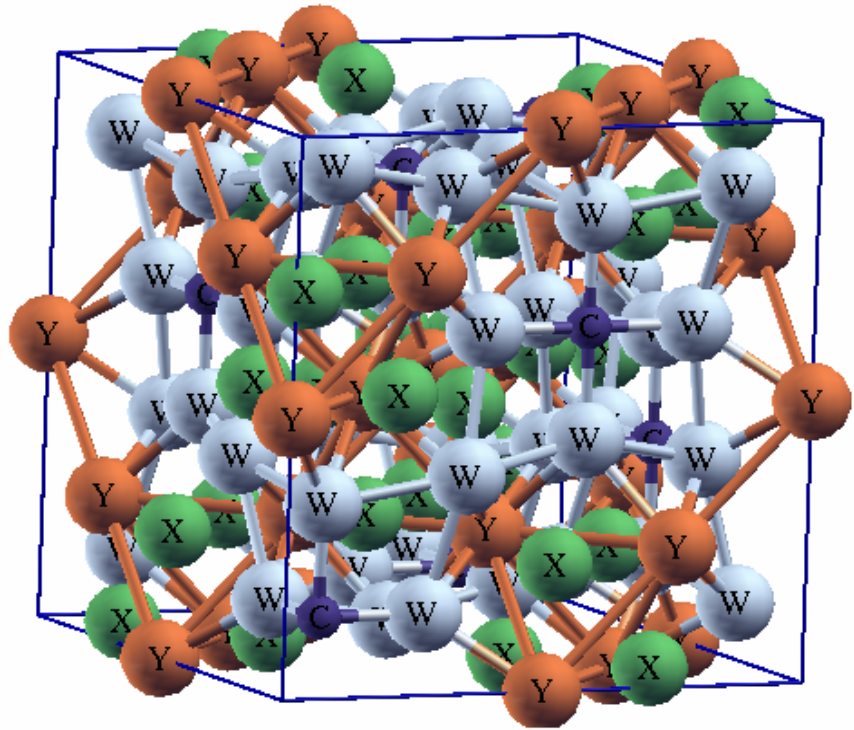

Fig. 1. The models of crystal structures for the ternary η carbides $M_3W_3C$ and $M_6W_6C$. Here, X=$M_1$ and Y=$M_2$ depict two different types of 3d – metal atoms placed in the 32(*e*) and the 16(*d*) positions.

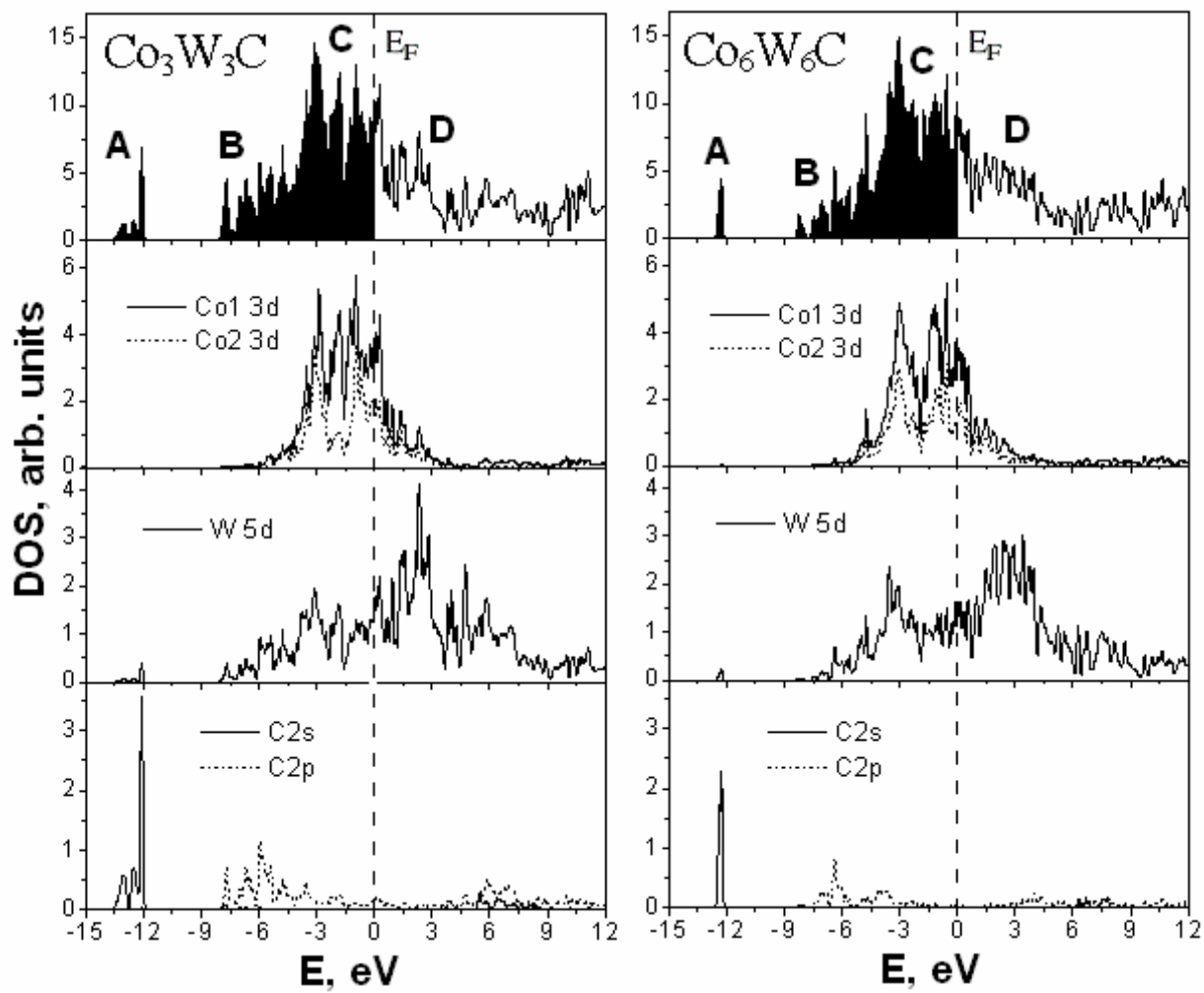

Fig. 2. Total (*upper panels*) and partial densities of states for the non-magnetic η carbides $Co_3W_3C$ and $Co_6W_6C$.

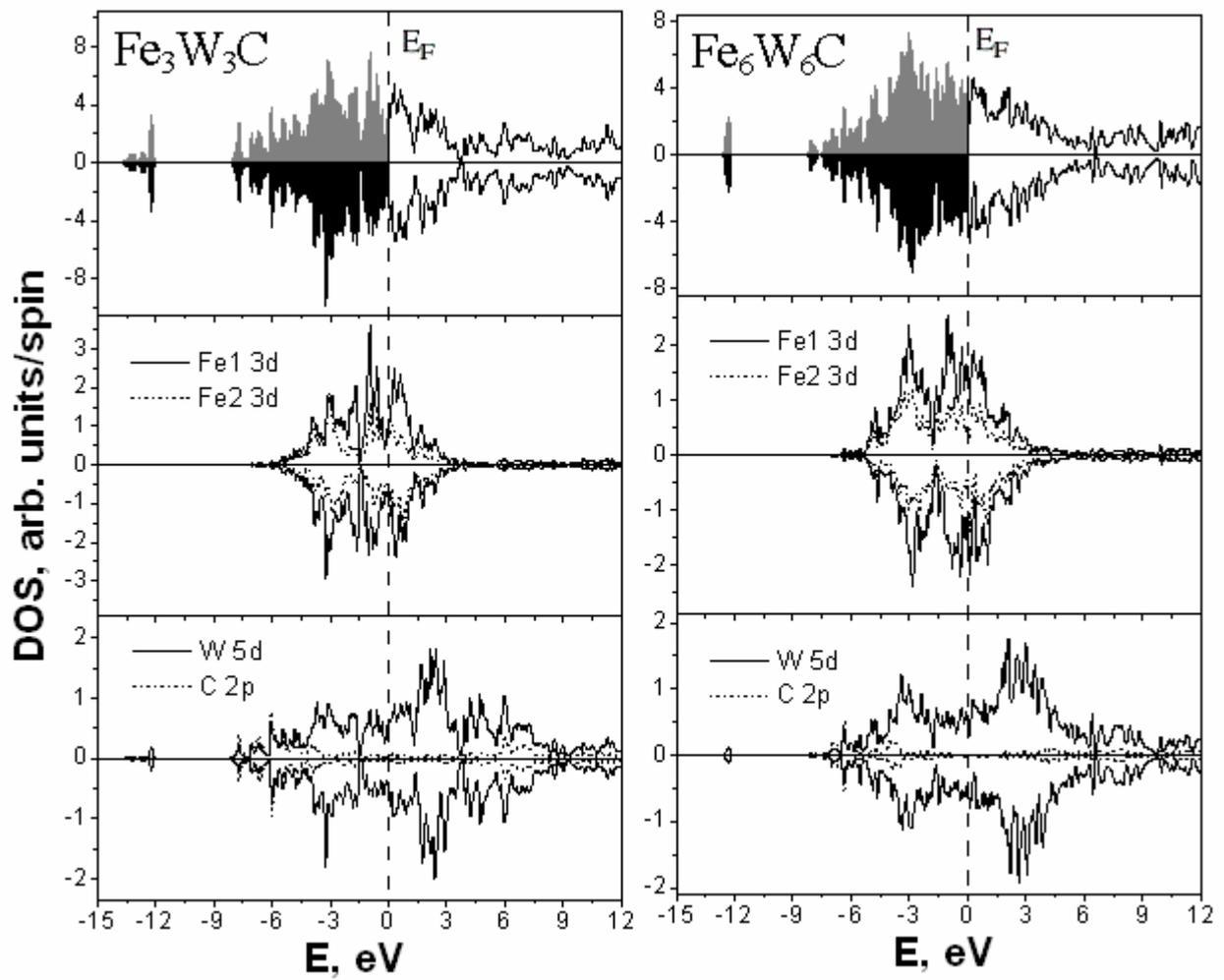

Fig. 3. Spin-resolved total (*upper panels*) and partial densities of states for the η carbides $Fe_3W_3C$ and $Fe_6W_6C$.